\documentclass[letterpaper,10pt]{article}
\usepackage{graphicx}
\usepackage{osameet3} %% use version 3 for proper copyright statement
\usepackage{mathtools}
\graphicspath{ {./images/} } 

\usepackage{amsmath,amssymb}
\usepackage[colorlinks=true,bookmarks=false,citecolor=blue,urlcolor=blue]{hyperref} %pdflatex

\begin{document}

\title{Quasi-distributed fiber sensing via perfect periodic Legendre codes}

\author{N Arbel$^1$, L Shiloh$^1$, N Levanon$^1$ and A Eyal$^1$}
\address{$^1$ School of EE, Tel-Aviv University, Tel-Aviv, Israel}
\email{nadavarbel@tauex.tau.ac.il}

\begin{abstract}

Long-range Rayleigh-based Distributed Acoustic Sensing (DAS) systems are often limited in their sensitivity and bandwidth. The former limitation is a result of the low backscattered power and poor \textit{dynamic-strain to optical-phase} transduction efficiency. The latter constraint results from the trade-off between range and scan-rate which limits the sampling interval to the longest delay in the sensing fiber. Quasi-DAS (Q-DAS) can yield enhanced sensitivity but may still suffer from low backscattered power and low scan-rate for long-haul, many-sensor, systems. In this work we study the use of Perfect Periodic Correlation codes for interrogating a long-haul Q-DAS system. It is shown that judicious choice of the code parameters allows order of magnitude increase in detection bandwidths and in the power reflected from each sensor.

\end{abstract}
\bigskip
%\keywords

%\submitto{\OFS}\maketitle
Submitted to: OFS - 2020

\section{Introduction}
% OFS vs Q-OFS -> where should we pick a QPFS 
Probably the most significant advantage of Fiber Optic Sensing (FOS) technology is that it provides a practical and cost effective way to deploy sensors along large distances or over wide areas. In a Distributed Fiber-Optic Sensor (D-FOS) the sensing is continuous along the entire fiber span whereas in a quasi-distributed fiber-optic sensing (QD-FOS) system there is typically a discrete number of fiber sensors which are connected via a single (or a pair of) transmission fiber(s). D-FOS is attractive due to the simple deployment and continuous coverage, however, in some applications QD-FOS may offer more adequate performance especially if high sensitivity is required. 

% Advantages of quasi distributed sensing: high SNR / long detection range / high sensitivity
The study of quasi-distributed fiber-optic sensors has been going on for about four decades. Applications include (but are not limited to) underwater acoustic sensing \cite{Kirkendall_2004}, structural health monitoring (SHM) \cite{tjin2002application},  Nuclear Power Plants (NPP) monitoring \cite{Ferdinand:97}, geophysical measurements \cite{liu2015sensing} and more. In all these applications the flexibility in optimizing the power which feeds each sensor and the transduction configuration have made them superior over their counterparts (D-FOS) in terms of SNR and sensitivity.  

% two configuration of Q-OFS: FBG and ladder
One popular example of QD-FOS is the Fiber Bragg Grating array (FBG array) \cite{kersey1997fiber}, where multiple FBG's are inscribed in the core of a single fiber and each one of them is used as a sensor. Another well established approach is the ladder or dual-bus configuration, where an array of optical sensors, such as interferometric sensors for example, are fed via an input-bus and their outputs are coupled to an output-bus \cite{brooks1985coherence, kersey1989multiplexed , kersey1988demonstration}. Half ladder configuration, where the same fiber is used as both the input and output fiber, are also of interest. Recently, Chen et. al. introduced an half-ladder configuration of strain sensors with very high sensitivity \cite{chen2017time}. 

% ladder configuration: advantages ... however it has two main problems: 1. poor SNR for array of sensors. 2. limited scan rate restricted by the fiber length.
Although QD-FOS are very attractive for applications which require very high sensitivity, their sensitivity degrades as the number of multiplexed sensors increases. Moreover, QD-FOS, like D-FOS, are restricted in their update rates due to the need to make the scan period longer than the maximum round-trip time in the sensing network. Hence, unless special techniques are used, the limit on the acoustic bandwidth of QD-FOS is directly related to the longest delay in the system. Namely, to prevent overlap of received signals from consecutive scans, it is required to satisfy $f_s<1/\tau_{max} \approx \ v/(2L)$, where $f_s$ is the update rate, $\tau_{max}$ is the longest delay in the system, $v$ is the speed of light in the fiber and $L$ is the total fiber length. Few approaches to overcome this limit have been studied. Using $N$ interrogation waveforms, each with a different center frequency (frequency division multiplexing), facilitated $N$-fold increase in the update rate~\cite{Bao2015,Iida2016}. Zhang et. al. ~\cite{zhang2017breaking} used random sampling (rather than the conventional uniform sampling) and detected MHz vibrations in a distributed sensor whose round-trip limit was \texttildelow 10kHz. These solutions are viable but the number of sensors is still limited due to the power splitting issue. In a ladder configuration, the power returning from a single sensing element is proportional to $1/N^2$ \cite{brooks1985coherence, chen2017time}, where $N$ is the number of elements in the configuration. Hence, scaling up the number of sensors is challenging. The use of special radar codes, rather than a single pulse, for interrogation of the measured system in order to improve the SNR and sensitivity, has already been demonstrated in many studies. For example, in a studies of a laser range finder (LRF) \cite{ arbel2016continuously }, distributed acoustic sensing (DAS)\cite{rosello2019distributed}, and more. However, in these studies the code duration was greater than $\tau_{max}$, improving the SNR but keeping the scan rate relatively low. This work studies the use of a code whose length is shorter than the fiber total length. Hence, achieving both very high scan rates and improved SNR and sensitivity.   

% explain what are PPA codes
A family of codes that offers optimal Peak to Sidelobe Ratio (PSR) are the Perfect Periodic Autocorrelation (PPA) codes. In theory, these codes have no sidelobes at all, namely, their inherent PSR is infinite. Moreover, as suggested by their name, the processing of raw data, which result from PPA code interrogation, involves periodic cross-correlation of the captured data. In contrast with non-periodic codes, this can be implemented directly using the highly efficient FFT and IFFT algorithms. The Legendre code, used in this work, is a member of the PPA-code family with length $N$ where $N$ can be any prime number that obeys $N = 4k-1$ and $k$ is an integer.

% Our solution
We present here for the first time a quasi-distributed sensing method that can detect acoustical signals at frequencies well beyond $1/\tau_{max}$ and can handle many sensing elements at relatively high SNR and sensitivity. We have experimentally demonstrated multiplexing of 3 sensing elements located at 1km, 26km and 76km. The scan rate was 12.5 kHz which is far greater than the standard limit associated with 76km fiber, i.e. \texttildelow 1.3 kHz. The implementation was based on coherent transmission and detection of Legendre sequence in a ladder configuration.

\section{Configuration and interrogation method}
\begin{figure}[ht] % h ->here t ->top
\includegraphics[width=16cm, height=3.6cm]{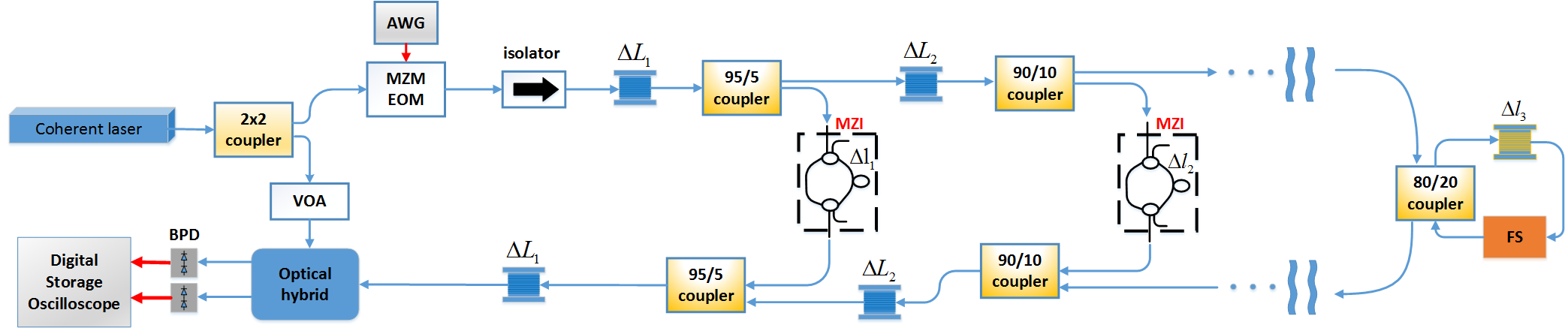}
\caption{ Quasi distributed optical fiber sensing setup. AWG: Arbitrary Waveform Generator. EOM: Electro-Optic Modulator. BPD: Balanced Photo-Detector. FS: Fiber Stretcher. VOA: Variable Optical Attenuator. MZI: Mach-Zehnder Interferometer. $\Delta L_1$ and $\Delta L_2$ are the lengths of the fiber delays to the 1st and 2nd ladder stages respectively. $\Delta l_1$ and $\Delta l_2$ are the MZI relative path differences. $\Delta l_3$ is the ring length.}
\label{setup}
\end{figure}

The system studied in this work was a coherent DAS interrogator connected to multiple discrete sensors in ladder configuration (Fig \ref{setup}).   

The output of an ultra-coherent laser source (linewidth \texttildelow 0.1kHz) was split between a reference arm and a sensing arm. The light at the reference arm was fed to the LO input of an optical hybrid. The light at the sensing arm was modulated by an Electro-Optic Modulator (EOM) and was launched into the input bus of the ladder. An Arbitrary Waveform Generator (AWG) was used to generate the desired PPA sequence and load it onto the optical carrier. The EOM output was connected to the ladder input through an optical isolator in order to prevent unwanted feedback mainly due to Rayleigh backscattering. Each of the ladder stages comprised a pair of identical tap couplers for input and output coupling and an unbalanced Mach-Zehnder interferometer (MZI) whose longer arm was the sensing element. To facilitate the characterization of the system the sensing elements were connected to fiber stretchers. The splitting ratios of the tap couplers were chosen according to their availability in the lab and with intention to equalize the power delivery to the sensors as much as possible. The light returning via the output-bus was mixed with the reference and detected by a pair of Balanced Receivers. A Digital Sampling Oscilloscope (DSO) was used for sampling and recording of the received signals.
The distances of the ladder stages from the interrogator were 1km, 26km and 76km. These distances were selected, again, according to availability of lab fiber-spools and with intention to demonstrate the ability of the system to operate over long distances. As described below, the number of sensors that the system can potentially support is $>100$. For convenience, each unbalanced MZI had a unique differential delay. This was done in order to instantly relate each pair of peaks to their ladder stage at the correlation output stage but was not fundamentally required by the method. 

To interrogate the sensors, the EOM launched a cyclic Legendre sequence into the input bus. The sequence period comprised of 4003 bits, each $20$ns long. Hence, the period duration, $\tau_{p}$, was \texttildelow 80\textmu sec which was much shorter than the longest delay in the system, $\tau_{max}$, associated with 76km. Copies of this sequence, with different time delays, reached the optical hybrid via its signal arm input. Each ladder element produced two closely delayed reflections except for the last element which, for simplification of the setup, was configured as a ring and produced \texttildelow4 reflections above the noise level.   

The detected raw data was cross-correlated with a digital copy of the same Legendre sequence yielding a distinct peak for each delayed replica of the transmitted sequence. By virtue of the PPA property, the peaks did not have any sidelobes and cross-talk was avoided. 

To allow scan rates higher than $1/{\tau_{max}}$ with no overlap of signals from different sensors, the code total duration was chosen with care. It was chosen such that $|mod(\tau_{i},\tau_{p})-mod(\tau_{j},\tau_{p})|>\delta$  $\forall$ i $\neq$ j, where $ \tau_i$ and $\tau_j$ are the roundtrip delays to the i-th and j-th ladder elements respectively (note that each ladder element comprises two peaks) and $\delta$ is predefined minimum separation between reflections.

\subsection{Experimental results}
\begin{figure}[ht] % h ->here t ->top

\centering
\includegraphics[width=7.5cm, height=4.4cm]{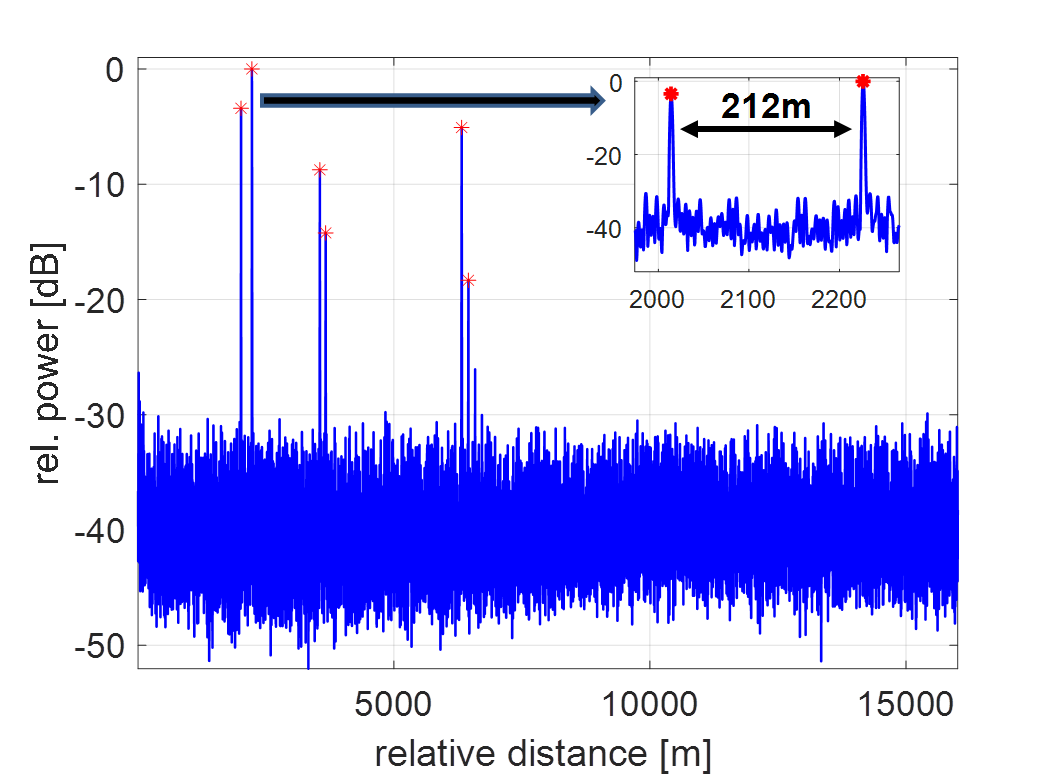}
\caption {The average power of the detected data after cross-correlation. The horizontal axis represents distance remainder in meters (namely, $mod(z,L_s)$ where $L_s$ is the length of the single period of the interrogation sequence). The vertical axis represents relative power in dB. A separation of 212m, 115m and 130m is evident between the first, second and third pairs of peaks.}
\label{mean_seg_power}

\end{figure}

The optical setup was comprised of three ladder elements located after 1km, 26km and 76km with respective path differences of 212m, 115m and 130m. A 4003 bits Legendre sequence with bit duration of 20ns was continuously launched into the fiber, having a total duration of \texttildelow80\textmu sec and a total length of \texttildelow 16km. A 15ms recording of raw data was cross-correlated with the digital reference resulting in \texttildelow 187 traces of the ladder profile. The average power of the traces is depicted in figure \ref{mean_seg_power}.

As can be seen in figure \ref{mean_seg_power}, a pair of correlation peaks with separation of 212m, which corresponds to the 1$^{st}$ MZI path difference, appears first to the left. A second pair of peaks with a separation of 115m is observed afterwards and corresponds to the 2$^{nd}$ MZI and finally a group of 4 peaks separated by an interval of 130m corresponds to the third ladder rung.  talk about the modulo  
The power difference between two peaks which correspond to the same MZI is a result of the FS insertion loss and possible polarization-related losses as the setup recorded only a single polarization component.

\begin{figure}[ht] % h ->here t ->top
\centering
\includegraphics[width=7cm, height=4.6cm]{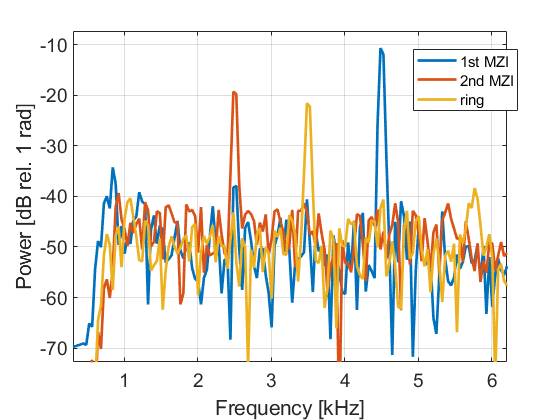}
\caption{ Spectra of detected signals from the 3 sensors. blue - the first MZI located at 1km, yellow - the second MZI located at 26km, red - the last ladder element, a ring, located at 76km.}
\label{detected_data_freq}
\end{figure}

In order to demonstrate the system's acoustic Bandwidth (BW), each sensor was excited with a Fiber-Stretcher (FS). The 1$^{st}$ FS operated at 4.5 kHz, the 2$^{nd}$ at 2.5 kHz and the 3$^{rd}$, located in the ring, operated at 3.5 kHz. A time interval of 25 msec was recorded and the differential phases between pairs of peaks which corresponds to the 3 sensors were extracted as functions of time. The spectra of the 3 differential phases are plotted figure \ref{detected_data_freq}. All signals had SNR greater than 22dB. In addition, no cross-talk was observed between the different sensing elements. The maximal detectable frequency stems from the sequence duration and was set in our experiment to 6.25 kHz.

\section{{Discussion and conclusions}}
In this work the use of a PPA code for interrogating a QD-FOS was demonstrated. The code period was much shorter than the longest delay in the sensing system. The number of sensors which can be implemented can be estimated from $\tau_p/\delta$ provided we can find parameters which satisfy  $|mod(\tau_{i},\tau_{p})-mod(\tau_{j},\tau_{p})|>\delta$  $\forall$ i $\neq$ j. Introducing design rules for obtaining valid parameters is beyond the scope of this paper. It is, however, rather easy to find valid parameters even with trial and error. As an example, it can be easily verified that 301 sensors, equally distributed along 40km, along with $\tau_p = 80\mu s$ and $\delta = 250ns$ results in non-overlapping responses. 

As in all ladder configurations, the power returning from an individual element is proportional to $1/N^2$, thus power considerations are critical to determine how many sensors can be supported. In this work the ladder was interrogated with a continuous waveform leading to a gain of $M$ (the code length) compared with single pulse interrogation.

In conclusion, the method provided significant increase in acoustic BW compared with conventional OFS methods. 
In addition, coding gain significantly increases the potential number of supported sensors. An experimental demonstration of 3 sensing elements, at a ranges of 1km, 26km and 76km was made. The system successfully detected 3 acoustic frequencies which exceeded the standard frequency limit related with setup longest delay.

\bibliographystyle{IEEEtran}
\bibliography{references1}
\end{document}